# Using the Relic Dark Energy hypothesis to investigate the physics of Cosmological expansion

A. W. Beckwith

Department of Physics and Texas Center for Superconductivity and Advanced
Materials at the University of Houston
Houston, Texas 77204-5005 USA

## ABSTRACT

We use a transplanckian dark energy hypothesis developed by Mersini et al to investigate cosmological expansion physics. Their parametric oscillator equation frequency can lead to a time dependent frequency; the dispersion relationship behavior mimics the Epstein functions used by Mersini et al in the initial phases of cosmological expansion. However, it is very difficult to meet the transplanckian assumptions of a vanishing of this same frequency with known physics at ultra high momentum values. Instead, we propose using a numerical algorithm; reconstructing the scale factor for cosmological expansion reconstructs more of the physics of the transplanckian hypothesis in transplanckian momentum values.

Correspondence: A. W. Beckwith:    projectbeckwith2@yahoo.com

# I. INTRODUCTION

We investigate whether using dark energy[1] from the tail-mode of ultra high momentum contributions of the universe leads to useful cosmological expansion models. We assert that the answer is yes, provided that we pick fitting scale factors for cosmological expansion.

The traditional Riedmann-Lemaitre-Robinson-Walker (FLRW) background includes small perturbations line element[2] with

$$d\ s^2 \equiv a^2(\eta) \cdot \{\ \} \tag{1}$$

Here the brackets contain scalar sectors — a traceless, transverse tensor $h_{ij}$ cc— that we associate with gravitational waves. We can use this tensor to define a quantity $\mu_T$ for each mode $k$ with a Fourier type of decomposition, which includes additional information via an assumed polarization tensor $p_{ij}^s(k)$. This will lead to an equation of state, similar to a time independent Schrödinger equation, due to a perturbed vacuum Einstein equation that we can write as[1,2]

$$\mu_T'' + \left(k^2 - \frac{a''}{a}\right)\mu_T = 0 \tag{2}$$

Making an analogy to time independent Schrödinger equation permits[2] us to write an effective potential, $U_T(\eta) \equiv a''/a$, and we me also use the analogy with respect to a parametric oscillator using the identification a time dependent frequency given by[2]

$$\omega_T \equiv k^2 - a''/a \tag{3}$$

We make a linkage to the transplanckian hypothesis by setting this as the square of the Mancini et al[1] derived frequency and the Epstein formula they used to obtain desired behavior of density functionals in a given ratio. They attempt to reconstruct a frequency range via modification of a hypothesis by Magueijo and Smolin[3] with respect to an alteration of the special relativity energy hypothesis advanced to fit cosmic ray data. However, that, plus a modification of Magueijo and Smolin[3] hypothesis, did not work out so well. Note that this is independent of the slow roll hypothesis needed for cosmological potential field systems.[4] We are referring to $\left|\frac{\partial^2 V}{\partial \phi^2}\right| << H^2$ where H is the Hubble expansion rate, which is a requirement of realistic inflation models.[4] Note that the slow roll requirement is for scalar fields dominant in the early phases of inflationary cosmology. In this situation, dynamics given by dominating gravitational wave perturbations lead to considerations of a momentum spread — and not the scalar field inflation model, which is useful in initial phases of cosmological inflation. Still, the break down of the dispersion relationship model as given an example will lead us to evaluate this problem via the numerical reconstruction procedure that we outline herein.

## II. USING MAQUEIJO AND SMOLIN EQUATIONS TO FIND ACCEPTABLE FREQUENCY BEHAVIOR

It is possible to make a derived dispersion relationship $a_M(k)$ that would match the Epstein function used by Mersini et al?[1] The original form of Magueijo and Smolin[3] energy expression is not appropriate for a dispersion relationship matching the Epstein function; we would have a violation of the slow roll requirement (from dispersion relationship potentials derived from Magueijo and Smolin[3] energy expression) that cosmologists use in inflationary expansion models. We followed Mersini et al[1] in their derivation of a transplanckian dark energy over total energy ratio being calculated via a Bogoliubov coefficient[5] in order to have a value less than ten to the minus 30 power that mimics the Epstein function[1,5] of special relativity. Linking our results with Magueijo and Smolin[3] energy expression appears to have insurmountable problems, one of which shows up in the beta coefficient in the denominator becoming too large. This is in tandem with Lemoine, Martin, and Uzan.[5,6] However, we assert that the transplanckian hypothesis is a useful means to winnow appropriate scale factor behavior — and should be viewed as a test-bed for finding models that appropriately characterize the expected expansion behavior physics when the universe evolves beyond today's environment.

## III. DESCRIPTION OF PROCEDURE USED TO OBTAIN ENERGY DENSITY RATIO.

What Mersini et al[1] did was to use ultra low dispersion relationship values for ultra high momentum values to obtain ultra-low energy values, which were, and still remain, allegedly frozen.[1] Using the Epstein function for frequency dispersion relationships, they found, a range of frequencies $\leq H_0$, where $H_0$ is the present Hubble rate of expansion. From there, they computed transplanckian dark energy modes, which are about 122 to 123 orders of magnitude smaller than the total energy of the universe assumed for their expansion model. Note that $\omega_K(k)$ refers to the dispersion relationship Mersini et al[1] derived; they changed a standard linear dispersion relationship to one that has a modified Epstein function with a peak value for frequency given when k = $k_C$; and, where, if we can set $k << k_C$, we have[1]

$$\omega_K^2(k) \approx k^2 \tag{4}$$

Thus, for low values of momentum we have a linear relationship for dispersion vs. momentum in low momentum situations. In addition we also have that[1]

$$\omega_K^2(k \gg k_C) \approx \exp(-k/k_C) \xrightarrow{k \to \infty} 0 \tag{5}$$

We also have a specific tail-mode energy region picked by

$$\omega_K^2(k_H) \equiv H_0^2 \tag{6}$$

to obtain $k_H$. We then have an energy calculation for the tail-modes[1]

$$\langle \rho_{TAIL} \rangle_K = \frac{1}{2 \cdot \pi^2} \cdot \int_{K_H}^{\infty} k \, dk \int \omega_K(k) \cdot d\omega_K \cdot |\beta_k|^2 \tag{7}$$

This is about 122 orders of magnitude smaller than

$$\langle \rho_{TOTAL} \rangle_K = \frac{1}{2 \cdot \pi^2} \cdot \int_0^{\infty} k \, dk \int \omega_K(k) \cdot d\omega_K \cdot |\beta_k|^2 \tag{8}$$

allowing us to write[1]

$$\frac{\langle \rho_{TAIL} \rangle_K}{\langle \rho_{TOTAL} \rangle_K} \approx \frac{k_H^2}{M_P^4} \cdot \omega_K^2(k_H) \approx \frac{H_0^2}{M_P^2} \approx 10^{-122} \tag{9}$$

Here, the tail-modes of energy are chosen as frozen[1] during any expansion of the universe. This is for energy modes for frequency regions $\omega_K^2(k) \leq H_0^2$ so that we have resulting tail-modes of energy obeying Eq. (9).

This Planck energy is the inverse of the Planck length defined by $l_P = \sqrt{\hbar \cdot G/c^3} \approx 10^{-44}$ cm, where G is the gravitational constant and c is the speed of light. Specifically, Magueijo and Smolin[3] state that $E_{PARTICLE} = E_P$ if and only if the rest mass of a particle obtains an infinite value. If we set $\hbar = c = 1$, we have $[M = M_P] = [E_P]$ as an upper bound. This upper bound with respect to particle energy is consistent with respect to four principles elucidated by Magueijo and Smolin,[3] which are:

i. Assume relativity of inertial frames: When gravitational effects can be neglected, all observers in free, inertial motions are equivalent. This means that there is no preferred state of motion.

ii. Assume an equivalence principle: Under the effect of gravity, freely falling observers are all equivalent to each other and are equivalent to inertial observers.

iii. A new principle is introduced: The observer independence of Planck energy. That is, that there exists an invariant energy scale which we shall take to be the Planck energy.

iv. There exists a correspondence principle: At energy scales much smaller than $E_P$, conventional special and general relativity are true.

That is, they hold to first order in the ratio of energy scales to $E_P$. We now can consider how can fashion these principles into predictions of energy values that we can use to obtain dispersion relationships. Magueijo and Smolin[3] obtained a modified relationship between energy and mass:

$$E_0 = \frac{m_0 \cdot c^2}{1 + \frac{m_0 \cdot c^2}{E_P}} \tag{10}$$

which, if $m = \gamma \cdot m_0$ and c set = 1, becomes:

$$E = \frac{m}{1 + \dfrac{m}{E_P}} \tag{11}$$

If we use Eq. (8) to obtain a dispersion relationship, after using $\hbar \equiv 1$, we would have the situation of

$$\frac{\langle \rho_{TAIL} \rangle_K}{\langle \rho_{TOTAL} \rangle_K} \approx 1 \neq 10^{-122} \tag{12}$$

This is primarily due to what the dispersion relationship $a$ would turn into, not to mention what $k_H$ would become. Next, we turned to a variation of the energy expression, with fit the Epstein function behavior for energy but which also violated the slow roll requirements of potential fields in cosmology. Needless to say we found it instructive to try to come up with an Ansatz, see discussion of Eq. (26), that appeared to give correct behavior to getting a ratio of Eq. (6). We found it useful to work with, instead with:

$$E = \frac{m}{\left(1 + \beta \cdot \dfrac{m}{E_P}\right)^{11}} \left(1 - \frac{m}{E_P}\right) \tag{13}$$

The factor of 11 was merely put in to help form a dispersion relationship approaching the Epstein function. Similarly, the $\left(1 - \dfrac{m}{E_P}\right)$ expression was included to get a tail-off behavior of a dispersion relationship. The only remote justification for this is that it gives us preferred numerical values we are seeking for the ratio of dark energy over total cosmological energy If $E_{PARTICLE} < E_P$ and $m = a \cdot k$, then $\dfrac{m}{E_P} \equiv \dfrac{k}{k_P} < 1$ permits a rewrite of Eq. (9) as (if $\beta \equiv 1000$):

$$\omega_M(k) = \frac{\alpha \cdot k}{\left(1 + \beta \frac{k}{k_P}\right)^{11}} \cdot \left(1 - \frac{k}{k_P}\right) \tag{14}$$

where we used $\hbar = c = 1$ and $[E] = [\hbar \cdot \omega] = [\omega_K(k)]$, which if $k \ll k_P$ will lead to the same result as spoken of with the modified Epstein function,[2] assuming that $|\alpha|^2 \cong 1$, so:

$$\omega_M^2(k) \approx k^2 \tag{15}$$

Furthermore, if $k - k_P - \varepsilon_{\text{+}}$, Eq. (14) will give us

$$\omega_M^2(k_P - \varepsilon_{\text{+}}) \cong \varepsilon_{\text{+}} \tag{16}$$

which, if $\omega_1(k) \equiv \omega_M(k)$, gives the values seen in Fig. 1

Note how the cut off value of momentum $k_P$ is due to $\left(1 - \frac{k}{k_P}\right)$ as a quantity in dispersion behavior leads to the results seen in Fig. 1.

**[Insert Fig. 1 about here]**

We can contrast this dispersion behavior with

$$\omega_1(k) = \frac{\alpha \cdot k}{\left(1 + \beta_1 \frac{k}{k_P}\right)^{11}} \cdot \exp\left(-\beta_2 \cdot \frac{k}{k_P}\right) \tag{17}$$

We set $\beta_1 \equiv 1$ and $\beta_2 \equiv 100$, leading to

$$\langle \rho_{TAIL} \rangle_M = \frac{1}{2 \cdot \pi^2} \cdot \int_{K_H}^{K_P} k dk \int \omega_M(k) \cdot d\omega_M \cdot |\beta_k|^2 \tag{18}$$

and

$$\langle \rho_{TOTAL} \rangle_M = \frac{1}{2 \cdot \pi^2} \int_0^{K_P} k\,dk \int \omega_M(k) \cdot d\omega_M \cdot |\beta_k|^2 \tag{19}$$

so we obtain [2] a frozen tail-mode energy vs. total energy ratio of

$$\frac{\langle \rho_{TAIL} \rangle_M}{\langle \rho_{TOTAL} \rangle_M} = \frac{\int_{K_H}^{K_P} k\,dk \int \omega_M(k) \cdot d\omega_M \cdot |\beta_k|^2}{\int_0^{K_P} k\,dk \int \omega_M(k) \cdot d\omega_M \cdot |\beta_k|^2} < 10^{-30} \text{ and } \neq 10^{-122} \tag{20}$$

when we are using $k_H \leq \frac{k_P}{2}$. Eq. (20) has a lower bound $\approx 10^{-122}$ as stated by Mersini [1] in Eq. (9) if we use $a_M(k_H) \approx H_0$. Detuning the sensitivity of this ratio to exact $k_H \div (M) \cdot k_P$ for any $M < 1$ is extremely important to the viability of our physical theory about how dark matter plays a role in inflationary cosmology.

## IV. UNABLE TO MATCH COSMIC RAY DATA, THE DISPERSION RELATIONSHIP WAS PHYSICALLY UNTENABLE

$\beta \equiv 1000$ in Eq. (14) was picked so $k_H$ could have a wide range of values. This permitted $\frac{\langle \rho_{TAIL} \rangle_M}{\langle \rho_{TOTAL} \rangle_M}$ to be bounded below by a value $\div 10^{-30}$ for $k_H \leq \frac{k_P}{2}$ in line with de-tuning the sensitivity of the ratio results if we use $\beta \equiv 1000$ in the Eq. (14) dispersion relationship. We obtain Mersini et al's[1] main result at the expense of not matching cosmic ray data.[1] We should note that Eq. (17) leads to a far broader dispersion curve width that also necessitated a far larger $k_H$ value, which was necessary to have the frequency $a_M(k_H) \approx H_0$ used by Mersini et al.[1] This in turn leads to a much bigger value for a lower bound for Eq. (20) than what would obtain numerically if we used Eq. (17)

for dispersion. Detuning the sensitivity of this ratio to be $k_H \backsim (M) \cdot k_P$ for any $M < 1$ is extremely important to the viability of our physical theory about how dark matter plays a role in inflationary cosmology. We find that this result is still not sufficient to match the cosmic ray problem[1] since Eq. (14) gives us:

$$\omega_M(k) \xrightarrow[K<<K_P]{} \frac{k}{\left(1+\beta_3 \frac{k}{k_P}\right)} \tag{21}$$

Note, $\beta_3 \cong 11 \cdot 10^{+3}$, whereas we would prefer to find $\beta_3 \cong 11 \cdot 10^{-10\ 7}$.

Can $\beta_3 \cong 11 \cdot 10^{-10}$ with a modified dispersion relationship based upon Magueijo and Smolin[3] hypothesis? The answer is no even after a modification of our dispersion relationship with $L = 2$, then[3] put in.

$$\omega_2(k) = \frac{\alpha \cdot k}{\left(1+\beta\left(\frac{k}{k_P}\right)^L\right)^{11}} \cdot \left(1-\left(\frac{k}{k_P}\right)^L\right) \tag{22}$$

However, even with a value of L=2 in Eq. (21) we obtained, for $\beta \equiv 2.25$ and $k_H \equiv \frac{k_P}{2}$

$$\frac{\langle \rho_{TAIL} \rangle_2}{\langle \rho_{TOTAL} \rangle_2} = \frac{\int_{K_H}^{K_P} k dk \int \omega_2(k) \cdot d\omega_2 \cdot |\beta_k|^2}{\int_0^{K_P} k dk \int \omega_2(k) \cdot d\omega_2 \cdot |\beta_k|^2} \leq 6.425 \cdot 10^{-3} \tag{23}$$

which has a very different lower bound than the behavior seen in Eq. (20). If we pick $\beta \equiv 10^{-10}$ as suggested by T. Jacobson[7] to try to solve the cosmic ray problem, we then find that Eq. (23) approaches unity. Appendix II shows us that we still could not match the beta coefficient values needed to solve the cosmic ray problem of special relativity.

# V. CONFRONTING THE NUMERICAL RECONSTRUCTION ALGORITHM NECESSITY TO RETRIEVE PHYSICS.

In the onset of this article, we noted the necessity of retrieving information with respect to

$$\omega_T \equiv k^2 - a''/a \cong \left(F^2(k) = (k^2 - \tilde{k}_1^2) \cdot V_0(x, x_0) + k^2 \cdot V_1(x - x_0) + k_1^2\right) \equiv \omega_K^2 \qquad (23a)$$

with

$$x = \frac{\tilde{k}}{k_P} \qquad (23b)$$

and

$$\tilde{k}_1 < k_P \qquad (23c)$$

and

$$V_0(x, x_0) = \frac{C}{1+e^X} + \frac{E \cdot e^X}{(1+e^X) \cdot (1+e^{X-X_o})} \qquad (23d)$$

and:

$$V_1(x - x_0) = -B \cdot \frac{e^X}{(1+e^{X-X_o})^2} \qquad (23e)$$

As we have that when $x = \dfrac{\tilde{k}}{k_P} \to x_0 \ll 1$

$$\omega_T \equiv k^2 - a''/a \xrightarrow[x \to \varepsilon^+]{} k^2 \qquad (24)$$

That is, we have that at low momentum values, the ratio of $a''/a$ becomes a vanishingly small contribution that grows with increasing time increments. So we can, using Fig. 1, set up qualitative bounds as to the behavior of the scale factor $a$ and use it to reconstruct useful physics, assuming that the transplanckian hypothesis is legitimate. Qualitatively, as the momentum gets larger, one sees

$$a''/a \to k^2 = \text{very large value} \tag{25}$$

That is, the scale factor expansion of the universe continues to grow, perhaps exponentially. No surprise here. The problem is in reconstructing what to expect in between these two momentum ranges. My preferred solution is first of all to make an Ansatz for the momentum values along the lines of

$$k \equiv k_{init} + c_1 \cdot k_{evol} \cdot \tau^A \tag{26}$$

Where the $A$ values could be varied as seen fit, from either positive or negative values, and then look at a numerical simulation along the lines of

$$\Delta^2 a_i = \frac{a_{i+1} - a_i}{\tau_{i+1} - \tau_i} - \frac{a_i - a_{i-1}}{\tau_i - \tau_{i-1}} \tag{27a}$$

We can use an initial starting point of

$$\Delta^2 a_1 = \frac{a_2 - a_1}{\tau_2 - \tau_1} - \frac{a_1 - a_{initial}}{\tau_1 - (\tau_{initial} - \tau^*)} \tag{27b}$$

This would lead to a numerical differential equation of the form

$$\Delta^2 a + \left(\omega_T(k(\tau)) - k^2\right) \cdot a \equiv 0 \tag{27c}$$

This is our situation, given what we know of the Epstein function behavior given in Fig. 1 definable as

$$\Delta^2 a - \left(\left|\omega_T(k(\tau)) - k^2\right|\right) \cdot a \equiv 0 \tag{27d}$$

As we observe $\omega_K^2 \equiv \omega_T \equiv k^2 - a''/a \xrightarrow[x \to \varepsilon^+]{} k^2$, we have the completely expected phenomena of a slowly expanding scale factor of $a$. What is of very significant import is determining the physics of the large hump in Fig. 1, that is, the inflection point where the net frequency turns from an increasing function of momentum to one where the frequency is decreasing asymptotically. I do believe that this has been insufficiently explored in the literature and bears significant impact upon solving necessary and sufficient conditions for the validity of the transplanckian hypothesis. Furthermore, note that this hump occurs well before the defining frequency of $\omega_K^2(k_H) \equiv H_0^2$. This is important as to the tail modes hypothesis of the transplanckian hypothesis, and would tie in with modeling how the momentum of our model varied with time $\tau$.

## VI. CONCLUSION

We found that the dispersion relationship given in Eq. (14) and its limiting behavior shown in Eq. (21) gives the lower bound behavior as noted in Eq. (21) for a wide range of possible $k_H : M \cdot k_P$ values if $M < 1$. This was, however, done for a physically unacceptably large $\beta \equiv 10^3$ value;[3,6,7] instead, we wanted[7] $\beta \equiv 10^{-10}$ in order to solve the cosmic ray problem.[3,6,7] We have thereby established that perhaps analytical criteria used to derive the behavior of the dispersion relationship is not necessarily the optimal way to extract physical information from the transplanckian hypothesis. Accordingly, in the last section, a necessary and sufficient set of conditions for a numerical simulation was proposed that would:

1. Give an idea of the relative importance of time variation for momentum in our cosmology models.

2. Lend itself to a numerical treatment to explain the maximum value of frequency as represented by Fig. 1.

3. Avoid, in particular, the violation of the slow roll condition for potential fields, which the analytical solution of this problem could not answer. As has been indicated, tying in the physics of the maximum value of the frequency obtained in the dispersion relationship would be very important in obtaining an explanation as to why the change from linearity occurred in dispersions relationships.

We will report our exploration of these issues in a future publication.

# APPENDIX I:
# THE BOGOLIUBOV FUNCTION USED

We followed Bastero-Gil and Mersini's[5] assumption of negligible deviations from a strictly thermal universe, and we proved it in our Bogoliubov coefficient calculation. This led to us picking the thermality coefficient[5] B to be quite small. In addition, the ratio of confocal times as given by $\left|\frac{\eta}{\eta_C}\right|$ had little impact upon Eq. (16) of the main text. Also, $x_0 = \frac{k}{k_P} \leq 1$. Therefore,

$$|\beta_k|^2 \equiv \frac{\sinh^2\left(\pi \cdot \frac{B}{2} \cdot \frac{1}{k} \cdot \left|\frac{\eta}{\eta_C}\right|\right) + \cos^2\left(\frac{\pi}{2} \cdot \sqrt{1 - 4 \cdot B \cdot e^{-X_o}}\right)}{\sinh^2\left(\pi \cdot (2-B) \cdot \frac{1}{k}\left|\frac{\eta}{\eta_C}\right|\right) - \sinh^2\left(\pi \cdot \frac{B}{2} \cdot \frac{1}{k} \cdot \left|\frac{\eta}{\eta_C}\right|\right)} \quad (1)$$

We derive this expression in Appendix II. In addition, we should note that Bastero-Gil[8] has a website that delineates the size of tail energy density from dark matter as $\rho_X \approx 10^{-122} M_P^4$, which is consistent with our findings that the Bogoliubov functions given by Eq. (1) may be often approximated by a constant with small effects on calculating the ratio of energy for the tail vs. total energy[2] given in Eq. (6) in the main text.

# APPENDIX II: DERIVING THE BOGOLIUBOV COEFFICIENT

## Part 1. Initial assumptions

We derive the Bogoliubov coefficient, which is used in Eq. (16). We refer to Bastero-Gil and Mersini's article[5] that has a Bogoliubov coefficient that takes into account a deviation function $\Gamma(k_0, B)$, which is a measure of deviation from thermality[5] in the spectrum of co moving frequency values $\Omega_n(k)$ over different momentum values. Note that $\eta$ is part of a scale factor $a(\eta) = |\eta_C/\eta|$ and $k = n/a(\eta)$ so that momentum $k \propto |\eta|$. Also, if we are working with the conformal case of $\varepsilon = 1/6$ appearing[5] in :

$$\Omega_n^2 = a^2(\eta) \cdot \omega_{NON-LIN}^2(k) - (1 - 6 \cdot \varepsilon) \cdot \frac{a''}{a} = a^2(\eta) \cdot \omega_{NON-LIN}^2(k) = a^2(\eta) \cdot F^2(k) \qquad (1)$$

then for small momentum

$$\omega_{NON-LIN}^2(\tilde{k}_0) \approx \tilde{k}_0^{\,2} \qquad (2)$$

if momentum $\tilde{k}_0 \ll k_P$, where we use the same sort of linear approximation used by Mersini et al[1], as specified for Eq. (17) of their article[1] if the Epstein function specified in Eq. (1) of the main text has a linear relationship. We write out a full treatment of the dispersion function $F(k)$ [5] since it permits a clean derivation of the Bogoliubov coefficient which has the deviation function $\Gamma(k_0, B)$. We begin with[5]:

$$|\beta_k|^2 \equiv |\beta_n|^2 = \frac{\sinh^2(2 \cdot \pi \cdot \hat{\Omega}_-) + \Gamma(k_0, B)}{\sinh^2(2 \cdot \pi \cdot \hat{\Omega}_+) - \sinh^2(2 \cdot \pi \cdot \hat{\Omega}_-)} \qquad (3)$$

where we get an appropriate value for the deviation function $\Gamma(k_0, B)$[4] based upon having the square of the dispersion function $F(k)$ obey Eqs. (1) and (2) for $\tilde{k}_0 \ll k_P$.

Note, $k_P$ is a maximum momentum value along the lines Magueijo and Smolin[3] suggested for an $E_P$ Plank energy value.

# Part 2. Deriving appropriate $\Gamma(k_0, B)$ deviation function values

We look at how Bastero- Gil and Mersini[5] obtained an appropriate $\Gamma(k_0, B)$ value. They wrote:

$$\Gamma(k_0, B) = \cosh^2\left(\frac{\pi}{2} \cdot \sqrt{4 \cdot B \cdot e^{-X_o} - 1}\right) \tag{4}$$

with

$$x_0 = \frac{\tilde{k}_0}{k_P} \ll 1 \tag{5}$$

and

$$F^2(k) = (k^2 - \tilde{k}_1^2) \cdot V_0(x, x_0) + k^2 \cdot V_1(x - x_0) + \tilde{k}_1^2 \tag{6}$$

where $\tilde{k}_1 < k_P$ and where $\tilde{k}_1$ is in the transplanckian regime but is much greater than $k_0$. We are determining what B should be in Eq. (16) of the main text provided that $F(k) \approx k$ as $x = \frac{\tilde{k}}{k_P} \to x_0$, which will lead to specific restraints we place upon $V_0(x, x_0)$ as well as $V_1(x - x_0)$ above. Following Bastero-Gil and Mersini,[5] we write:

$$V_0(x, x_0) = \frac{C}{1 + e^X} + \frac{E \cdot e^X}{(1 + e^X) \cdot (1 + e^{X - X_o})} \tag{7}$$

and:

$$V_1(x - x_0) = -B \cdot \frac{e^X}{(1 + e^{X - X_o})^2} \tag{8}$$

When $x = \dfrac{\tilde{k}}{k_P} \to x_0 \ll 1$ we get [1,5]

$$F^2(k_0) \equiv \omega^2_{NON-LIN}(k_0) \cong -k_1^2 \cdot \left(1 - \dfrac{c}{2} - \dfrac{E}{4}\right) + k_0^2 \cdot \left(\dfrac{c}{2} + \dfrac{E}{4} - \dfrac{B}{4}\right) \cong k_0^2 \qquad (9)$$

which implies $0 < B \approx \varepsilon_+ \ll 1$. Then, we obtain:

$$\Gamma(k_0, B \cong \varepsilon_+) \cong \cosh^2\left(\left(\dfrac{\pi}{2} + \varepsilon_+\right) \cdot i\right) \approx \varepsilon_+ \ll 1 \qquad (10)$$

and

$$|\beta_k|^2 \equiv |\beta_n|^2 \cong \dfrac{\sinh^2(2 \cdot \pi \cdot \hat{\Omega}_-) + \varepsilon_+}{\sinh^2(2 \cdot \pi \cdot \hat{\Omega}_+) - \sinh^2(2 \cdot \pi \cdot \hat{\Omega}_-)} \qquad (11)$$

## Part 3. Finding appropriate $\hat{\Omega}_+$ and $\hat{\Omega}_-$ values

We define, following Bastero-Gil and Mersini[5]

$$\hat{\Omega}_\pm = \dfrac{1}{2} \cdot \left(\hat{\Omega}_{OUT} \pm \hat{\Omega}_{IN}\right) \qquad (12)$$

where we have that

$$\Omega^{OUT} = \xrightarrow{\eta \to \infty} \Omega_n(\eta \equiv \infty) \qquad (13)$$

and

$$\Omega^{IN} = \xrightarrow{\eta \to -\infty} \Omega_n(\eta \equiv -\infty) \qquad (14)$$

whereas we have that

$$\hat{\Omega}_{\tilde{k}} = \dfrac{\Omega_{\tilde{k}}}{n} \qquad (15)$$

where $\tilde{k}$ denotes either out or in. Also:

$$\Omega^{OUT} \cong \Omega^{IN} \cong 1 \qquad (16)$$

which lead to:

$$\hat{\Omega}_+ \cong (1-\frac{B}{2})\cdot\frac{1}{n} = (1-\frac{B}{2})\cdot\frac{1}{k}\cdot\left|\frac{\eta}{\eta_C}\right| \cong \frac{1}{k}\cdot\left|\frac{\eta}{\eta_C}\right| \qquad (17)$$

as well as

$$\hat{\Omega}_- \cong \frac{B}{2}\cdot\frac{1}{n} \cong 0 \qquad (18)$$

# APPENDIX ENTRY III.
# HOW EQUATION 20 OF TEXT CHANGES FOR VARYING $\beta$ VALUES AND DIFFERENT DISPERSION RELATIONSHIPS.

Starting with equation 21 of the main text.

If $\beta = 1.05$ and L = ½, $\left(\dfrac{k}{k_P}\right) \to \sqrt{\dfrac{k}{k_P}}$, then $\dfrac{\langle \rho_{TAIL} \rangle_M}{\langle \rho_{TOTAL} \rangle_M} \cong .371$

If $\beta = 1.05$ and L=1, $\left(\dfrac{k}{k_P}\right) \to \left(\dfrac{k}{k_P}\right)$, then $\dfrac{\langle \rho_{TAIL} \rangle_M}{\langle \rho_{TOTAL} \rangle_M} \cong .263$

If $\beta = 1.05$ and L=2, $\left(\dfrac{k}{k_P}\right) \to \left(\dfrac{k}{k_P}\right)^2$, then $\dfrac{\langle \rho_{TAIL} \rangle_M}{\langle \rho_{TOTAL} \rangle_M} \cong .115$

If $\beta = 10.5$ and L = ½, $\left(\dfrac{k}{k_P}\right) \to \sqrt{\dfrac{k}{k_P}}$, then $\dfrac{\langle \rho_{TAIL} \rangle_M}{\langle \rho_{TOTAL} \rangle_M} \cong 1.935 \cdot 10^{-5}$

If $\beta = 10.5$ and L=1, $\left(\dfrac{k}{k_P}\right) \to \left(\dfrac{k}{k_P}\right)$, then $\dfrac{\langle \rho_{TAIL} \rangle_M}{\langle \rho_{TOTAL} \rangle_M} \cong 7.347 \cdot 10^{-6}$

If $\beta = 10.5$ and L=2, $\left(\dfrac{k}{k_P}\right) \to \left(\dfrac{k}{k_P}\right)^2$, then $\dfrac{\langle \rho_{TAIL} \rangle_M}{\langle \rho_{TOTAL} \rangle_M} \cong 6.7448 \cdot 10^{-8}$

We need $\beta \cong 10^{-10}$ with $\dfrac{\langle \rho_{TAIL} \rangle_M}{\langle \rho_{TOTAL} \rangle_M} \leq 10^{-30}$ to get our results via this transplanckian model to be consistent with physically verifiable solutions to the cosmic ray problem.

# FIGURE CAPTIONS

Fig. 1    Graph of dispersion relationship $a_M(k)$ against momentum. This gives the desired behavior in line with the transplanckian dark energy hypothesis. However, $\beta \equiv 10^3$!

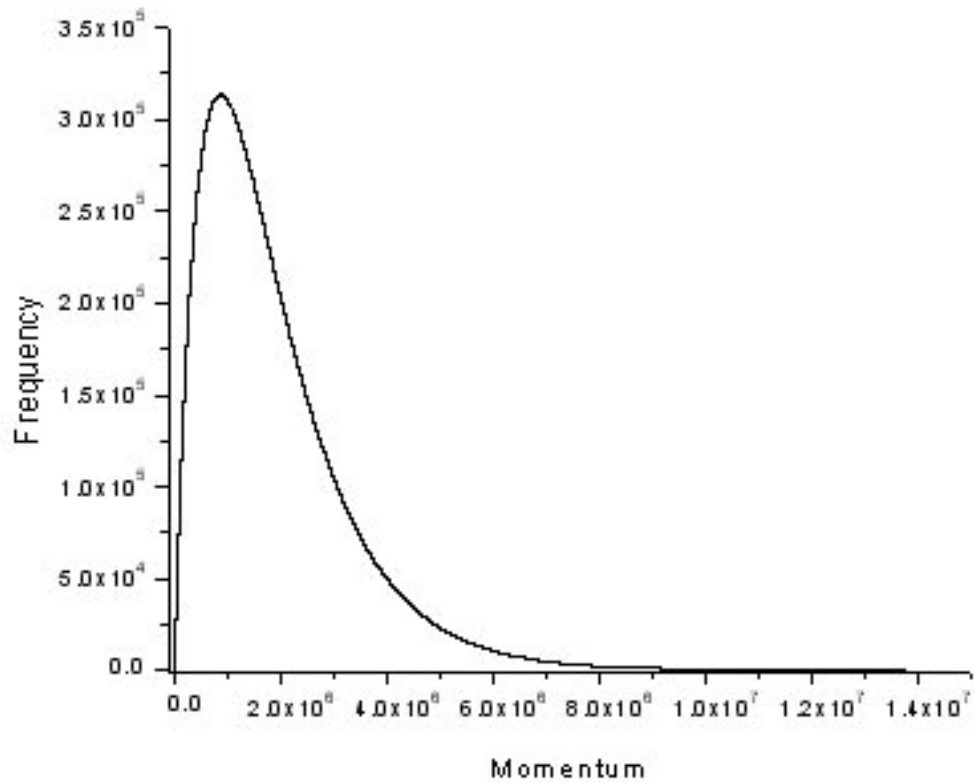

**Figure 1**
**Beckwith**